\documentclass[pra,a4paper,preprint,showpacs,superscriptaddress]{revtex4-2}
\usepackage{amsmath,bbm, bm}
\usepackage{amsfonts}
\usepackage{graphicx}
\usepackage{longtable} 
\newcommand{\be}{\begin{equation}}
\newcommand{\ee}{\end{equation}}

\newcommand{\ba}[1]{\left(\begin{array}{#1}}
\newcommand{\ea}{\end{array}\right)}
\begin{document}
	\title{Canonical structures of $A$ and $B$ forms} 
	\author{Sudha} 
	\email{tthdrs@gmail.com}
	\affiliation{Department of Physics, Kuvempu University, Shankaraghatta, Shimoga-577 451, India.}
	\affiliation{Inspire Institute Inc., Alexandria, Virginia, 22303, USA.}
	\author{B.N. Karthik}
	\affiliation{Department of Physics, Bangalore University, 
		Bangalore-560 056, India}
	\author{A. R. Usha Devi}
	\affiliation{Department of Physics, Bangalore University, 
		Bangalore-560 056, India} 
	\affiliation{Inspire Institute Inc., Alexandria, Virginia, 22303, USA.}
	\author{A. K. Rajagopal} 
	\affiliation{Inspire Institute Inc., Alexandria, Virginia, 22303, USA.}
	\date{\today}
	\begin{abstract} 
		In their seminal paper (Phys. Rev.121, 920 (1961)) Sudarshan, Mathews and Rau investigated  properties of the dynamical $A$ and $B$ maps acting on  $n$  dimensional quantum systems. Nature of the dynamical maps in open quantum system evolutions has attracted great deal of attention in the later years. However, the novel paper on the $A$ and $B$ dynamical maps has not received its due attention. In this tutorial article we review the properties of $A$ and $B$ forms associated with the dynamics of  finite dimensional quantum systems. In particular we investigate a  canonical structure associated with the $A$ form and establish its equivalence with the associated $B$ form. We show that the canonical structure of the $A$ form captures the completely positive (not completely positive) nature of  the dynamics in a succinct manner. This feature is illustrated through  physical examples of qubit channels. 
	\end{abstract}
	\maketitle
	\section{Introduction} 
	
	 The conceptual formulation of dynamical $A$ and $B$ forms was pioneered by Sudarshan and coworkers 60 years ago~\cite{ECGS1,Jordan1}. The $A$ and $B$ matrices play an important role to identify if the open system dynamics of finite dimensional quantum systems is completely positive or not~\cite{NC,Breuer,simon,Vinayak}. In this article we study the dynamical $A$ and $B$ forms in detail. In particular we investigate the canonical structure of the $A$ form and its properties. We show that there is a one-to-one connection between the canonical $A$ form and the $B$ form. We also construct the canonical $A$ form associated with some important physical examples of qubit channels~\cite{NC}.

In Sec.~II we introduce the $A$ and $B$ forms and discuss their properties~\cite{ECGS1,Jordan1}.  We present the canonical structure of the $A$ form and establish its equivalence with the $B$ form in Sec.~III.  The canonical structure of the $A$ form is explicitly constructed for several qubit channels in Sec.~IV. Concluding remarks are given in Sec~V.

\section{Properties of $A$ and $B$ maps} 

Consider a $n$ dimensional Hilbert Space $\mathcal{H}_n$.  State of a quantum system is described by a density matrix $\rho\in \mathcal{H}_n$, defining properties of which are given by  
\begin{enumerate}
	\item Hermiticity:  $\rho^\dag=\rho$. 
	\item Unit trace:  ${\rm Tr}\,\rho=1$. 
	\item Positivity: $\langle\,\psi\vert \rho\vert\psi\rangle\geq 0\,$\ for all \ \ $\vert\psi\rangle\in \mathcal{H}.$
\end{enumerate}
For a qubit (two-level quantum system) we have 
\begin{equation}
\label{q1}
\rho=\frac{1}{2}\,\left(I +\vec{\sigma}\cdot\, \vec{p}\right)
\end{equation}
where $I$ denotes $2\times 2$ identity matrix, $\vec{\sigma}=(\sigma_1,\sigma_2,\sigma_3)$ are  Pauli matrices: 
\begin{equation}
\sigma_1=\left(\begin{array}{cc} 0  & 1 \\ 
	1 & 0  \end{array}\right), \ \
\sigma_2=\left(\begin{array}{cc} 0  & -i \\ 
i & 0  \end{array}\right), \ \
\sigma_3=\left(\begin{array}{cc} 1  & 0 \\ 
0 & -1  \end{array}\right), \ \
\end{equation}
 and the real qubit state parameters given by  $\vec{p}=(p_1,p_2,p_3)$  satisfy the condition $\vert\vec{p}\vert=\sqrt{p_1^2+p_2^2+p_3^2}\leq 1$. Thus, the state space of a qubit corresponds to a unit ball in $\mathbb{R}^3$. The vector $\vec{p}$  is called the Bloch vector.    

Now we consider a linear map $A$ transforming density matrices in $\mathcal{H}_n$: 
 \begin{eqnarray}
 \label{defA}
 A:&&\,\rho_i\rightarrow\rho_f = A(\rho_i),  \nonumber \\ 
 \left(\rho_f\right)_{r's'}&=&\left(A\,\rho_i\right)_{r's'}=\sum_{r,s=1}^n\, A_{r's';rs}\, \left(\rho_i\right)_{rs}, \ \ \  r',s'=1,2,\ldots, n.
 \end{eqnarray}    
 \begin{itemize}
 \item The $n^2\times n^2$ matrix $A$ is called a trace-preserving positive map if, for every input density matrix, the
output $\rho_f = A(\rho_i)$ is also a legitimate density matrix~\cite{ECGS1,Jordan1}.  
\end{itemize}

Writing the qubit density matrix (\ref{q1})  explicitly  (in the standard basis $\vert 0\rangle=(1,0)^T$, $\vert 1\rangle=(0,1)^T$) as 
\begin{equation}
\label{qrho}
\rho=\left(\begin{array}{cc} \rho_{00}  & \rho_{01} \\ 
\rho_{10} & \rho_{11}  \end{array}\right)=\frac{1}{2}\left(\begin{array}{cc} 1+p_3 & p_1-i\, p_2 \\ 
                                 p_1+ip_2 & 1-p_3  \end{array}\right)
\end{equation}
one may identify the action $A:\ \rho_i\rightarrow\rho_f = A(\rho_i)$ (see (\ref{defA})) as follows: 
\begin{eqnarray}
\label{qfAqi}
\left(\begin{array}{c} \left(\rho_f\right)_{00} \\ \left(\rho_f\right)_{01}\\ \left(\rho_f\right)_{10}  \\ \left(\rho_f\right)_{11} \end{array}\right) &=& 
\left(\begin{array}{cccc}  A_{00;00} & A_{00;01} & A_{00;10} & A_{00;11}  \\  
                           A_{01;00} & A_{01;01} & A_{01;10} & A_{01;11}  \\
                           A_{10;00} & A_{10;01} & A_{10;10} & A_{10;11}  \\
                           A_{11;00} & A_{11;01} & A_{11;10} & A_{11;11}  
                           \end{array}\right)   
\left(\begin{array}{c} \left(\rho_i\right)_{00} \\ \left(\rho_i\right)_{01}\\ \left(\rho_i\right)_{10}  \\ \left(\rho_i\right)_{11} \end{array}\right) \nonumber \\ 
{\rm or} \ \ \  
\left(\begin{array}{c} 1+\left(p_f\right)_{3} \\ \left(p_f\right)_{1}-i \left(p_f\right)_{2}\\ \left(p_f\right)_{1}+i \left(p_f\right)_{2}  \\ 1-\left(p_f\right)_{3} \end{array}\right) &=& 
\left(\begin{array}{cccc}  A_{00;00} & A_{00;01} & A_{00;10} & A_{00;11}  \\  
A_{01;00} & A_{01;01} & A_{01;10} & A_{01;11}  \\
A_{10;00} & A_{10;01} & A_{10;10} & A_{10;11}  \\
A_{11;00} & A_{11;01} & A_{11;10} & A_{11;11}  
\end{array}\right)   
\left(\begin{array}{c} 1+\left(p_i\right)_{3} \\ \left(p_i\right)_{1}-i \left(p_i\right)_{2}\\ \left(p_i\right)_{1}+i \left(p_i\right)_{2}  \\ 1-\left(p_i\right)_{3} \end{array}\right). 
\end{eqnarray}

\begin{itemize}
	\item Unitary dynamics $\rho_f=U\,\rho_i\,U^\dag$ defines a trace-preserving positive map 
\begin{eqnarray}
\label{defAU}
A_U:&&\,\rho_i\rightarrow\rho_f = A_U(\rho_i),  \nonumber \\ 
\left(\rho_f\right)_{r's'}&=&\sum_{r,s=1}^n\, \left(U\otimes U^*\right)_{r's';rs}\, \left(\rho_i\right)_{rs}, \ \ \  r',s'=1,2,\ldots, n.
\end{eqnarray}    
\item Matrix transposition given by   
\begin{eqnarray}
\label{defAUT}
A_T:&&\rho_i\rightarrow\rho_f = A_T(\rho_i)=\rho_i^T  
\end{eqnarray}     
 is an example of  trace-preserving positive map. 
\end{itemize}
Given a positive map $A$ on $\mathcal{H}_n$  one may extend it to the map $A\otimes \mathbbm{I}_d$ acting on the tensor product space  $\mathcal{H}_n\otimes \mathcal{H}_d$  of a composite
$n\times d$ system, where $ \mathbbm{I}_d$ denotes identity map on $\mathcal{H}_d$. Then the positive map $A$ acts only on the  $n$-dimensional  subsystem of the composite state. If the $A\otimes \mathbbm{I}_d$ transforms a composite density matrix into a physical state for any $d$, then the map $A$ is said to be {\em completely positive} (CP). Otherwise, it is called {\em not-completely positive} (NCP).  Thus  a positive map represents a bonafide state to state transformation  if it is  completely positive.

Under the action of a map $A:\rho_i\Longrightarrow \rho_f=A(\rho_i)$, preservation of hermiticity i.e.,  $(\rho_f)^*_{s'r'}=(\rho_f)_{r's'}$ and the unit trace condition $\sum_{r'=1}^n\,\left(\rho_f\right)_{r'r'}=1$  result in the following constraints  on the elements of the $n^2\times n^2$ process matrix $A$:   
\begin{eqnarray}
\label{A1}
\left(\rho_f\right)_{r's'}&=&\left(\rho_f\right)^*_{s'r'} \Longrightarrow \sum_{r,s=1}^n\, A^*_{s'r';sr}\, \left(\rho_i\right)^*_{sr}=\sum_{r,s=1}^n\, A_{r's';rs}\, \left(\rho_i\right)_{rs} \nonumber \\ 
\Longrightarrow &&   \sum_{r,s=1}^n\, A^*_{s'r';sr}\, \left(\rho_i\right)_{rs} = \sum_{r,s=1}^n\, A_{r's';rs}\, \left(\rho_i\right)_{rs} \nonumber \\
\Longrightarrow &&  A^*_{r's';rs} = A_{s'r';sr}
\end{eqnarray}
and 
\begin{eqnarray}
\label{A2}
\sum_{r'=1}^n\left(\rho_f\right)_{r'r'}&=&1\Longrightarrow  \sum_{r',r,s}\, A^*_{r'r';rs}\, \left(\rho_i\right)_{rs}  \nonumber \\ 
\Longrightarrow &&   \sum_{r'=1}^n\, A_{r'r';rs}=\delta_{rs}  
\end{eqnarray}
where $\delta_{rs}$ denotes Kronecker delta symbol. 

A realigned process matrix $B$ was defined as~\cite{ECGS1,Jordan1} 
\begin{equation}
\label{abdef}
B_{r'r;s's}=A_{r's';rs}. 
\end{equation}  
so that the hermiticity and unit trace conditions (\ref{A1}), (\ref{A2}) on the $A$-form can be expressed as  
\begin{eqnarray}
\label{B1}
B_{r'r;s's}&=&B^*_{s's;r'r} \\ 
\sum_{r'=1}^n B_{r'r;r's}=\delta_{rs} && \Longrightarrow\ \  {\rm Tr}\,B=n.
\end{eqnarray}
Thus a physically valid  $A$-form  requires that the corresponding realigned matrix $B$ (see (\ref{abdef}))  is a $n^2\times n^2$ hermitian matrix  with trace $n$. 

Furthermore, positivity of the density matrix $\rho_f=A(\rho_i)\geq 0$ leads to the following constraints on the elements of $A$ and $B$ respectively~\cite{ECGS1}: 
\begin{eqnarray}
\sum_{r,s,r',s'}\,x^*_r\,x_s\, A_{rs;r's'}\, y_{r'}y_{s'}^*\geq 0, \nonumber \\ 
\sum_{r,s,r',s'}\, x^*_r\,y_{r'}\, B_{rr';ss'}\,x_s\, y_{s'}^*\geq 0. 
\end{eqnarray}
In other words, positivity $\rho_f=A(\rho_i)\geq 0$ of the density matrix requires that $B\geq 0.$  

It is pertinent to point out that    the $B$-form is represented by a hermitian matrix whereas $A$ is not; positivity of the $B$ matrix highlights that the output density matrix is legitimate.  For this reason  Sudarshan, Mathews and Rau~\cite{ECGS1} highlighted that  {\em the matrix $B$ incorporates the kinematical restrictions on the dynamical law in a succint fashion; we shall call $B$ the dynamical matrix}.  The $A$-form was used in Ref.~\cite{ECGS1} to define a linear map from input to output density operators (where the elements of the input and outpur density matrices are arranged in the form of  $n^2$ component  columns). Beyond this initial definition, the $A$ matrix was not recognized to have any clear role.  Our focus here is to unravel the $A$-form to its full potential. We show in the next section  that  the $A$ matrix introduced in the Sudarshan-Mathew-Rau paper exhibits an elegant canonical structure and it reveals itself as a powerful tool in capturing all the dynamical features reflected by the corresponding  $B$-form~\cite{urs}. 

%%%%%%%%%%%%%%%%%%%%%%%%%%%%%%%%%%%%%%%%%%%%%
\section{Canonical structure of the $A$-form}

Consider an orthonormal set $$\{T_\mu, \mu=0,1,2,\ldots, n^2-1\}$$ of $n\times n$  matrices satisfying  
\begin{equation}
\label{tor} 
{\rm Tr}[T_\mu^\dag\, T_\nu]=\delta_{\mu,\nu}.
\end{equation} 
We then construct a basis set 
\begin{equation}
\{T_\mu\otimes T_\nu^* , \ \  \mu,\nu=0,1,\ldots , n^2-1\}
\end{equation} 
of  $n^2\times n^2$ matrices and express the  $A$ matrix (see (\ref{defA})) as follows:      
\begin{equation}
\label{smalla1}
A=\sum_{\mu,\nu=0}^{n^2-1} \, a_{\mu\nu}\, T_\mu\otimes T_\nu^* 
\end{equation} 
where the expansion coefficients $a_{\mu\nu}$ are given by 
\begin{equation}
\label{asmall}
a_{\mu\nu}={\rm Tr}[A(T^\dag_\mu\otimes T^T_\nu)],\ \ \mu, \nu=0,1,\ldots , n^2-1.
\end{equation} 
The matrix elements $A_{r's';rs}$  of the $A$ matrix are then given by   
\begin{equation}
\label{a2}
A_{r's';rs}=\sum_{\mu,\nu=0}^{n^2}\, a_{\mu\nu}\, [T_\mu]_{r'r}\, [T^*_\nu]_{s's}.   
\end{equation} 
Let us examine the hermiticity preserving condition (\ref{A1}) on the expansion coefficients  $a_{\alpha\beta}$:  
\begin{equation}
\label{aher}
A^*_{r's';rs} = A_{s'r';sr} \Longrightarrow  
a^*_{\mu\nu}=a_{\nu\mu} 
\end{equation} 
In other words the coefficients $a_{\mu\nu}, \mu,\nu=0,1,\ldots, n^2-1$ constitute a $n^2\times n^2$ hermitian matrix, which we denote by  $\mathcal{A}$. 

Let $\mathcal{U}$ be a unitary matrix which diagonalizes $\mathcal{A}$ i.e., 
\begin{equation}
\label{uaudag}
\mathcal{U}\, \mathcal{A}\, \mathcal{U}^\dag=\mathcal{A}_0=\left(\begin{array}{lllll}\lambda_0 & 0 & \ldots & \ldots &  0 \\  
                                                                                     0 & \lambda_1 & \ldots & \ldots &0 \\ 
                                                                                     \vdots & \vdots & \ddots &  & \vdots \\ 
                                                                                     0 & 0 & \ldots & \ldots & \lambda_{n^2-1}   \end{array}\right)
\end{equation}
where $\lambda_\mu, 0\leq \mu\leq n^2-1$ denote the eigenvalues of $\mathcal{A}$. Thus we obtain        
\begin{eqnarray}
\label{ael}
a_{\mu\nu}&=&\left(\mathcal{U}^\dag\, \mathcal{A}_0\, \mathcal{U}\right)_{\mu\nu}\nonumber \\ 
&=&\sum_{\alpha}\lambda_\alpha\,  u^{*}_{\alpha\mu}\, \, u_{\alpha\nu}.
\end{eqnarray}
Substituting (\ref{ael}) in (\ref{smalla1}) we obtain the following canonical structure of the $A$ matrix:  
\begin{eqnarray}
\label{canA}
A&=&\sum_{\mu,\nu,\alpha}\,\lambda_\alpha\, u^*_{\alpha\mu}u_{\alpha\nu}\, \left(T_\mu\otimes T_\nu^*\right) \nonumber \\
&=& \sum_\alpha\, \lambda_\alpha\,\left( C_\alpha\otimes C_\alpha^*\right),
\end{eqnarray} 
where we have denoted 
\begin{equation}
C_\alpha=\sum_{\mu=0}^{n^2-1}\,  u^*_{\alpha\mu}\, T_\mu.
\end{equation}

\begin{itemize}
\item Using (\ref{canA}) we can express the matrix elements of $A$ as    
\begin{equation}
\label{canA2}
A_{r's';rs}=\sum_{\alpha=0}^{n^2-1}\, \lambda_\alpha\, (C_\alpha)_{r'r}\,(C_\alpha^*)_{s's}.
\end{equation}
Substituting (\ref{canA2}) in (\ref{defA}) and simplifying, we obtain the following elegant structure for the action of the linear $A$-map on the column vector consisting of the elements of the  input density matrix $\rho_i$:   
\begin{eqnarray}
\label{ele}
(\rho_f)_{r's'}&=&\sum_{r,s}\, A_{r's';rs}\, (\rho_i)_{rs} \nonumber \\
&=&  \sum_{r,s,\alpha} \, \lambda_\alpha\, (C_\alpha)_{r'r}\,(C^*_\alpha)_{s's}\,(\rho_i)_{rs}     \nonumber \\
&=& \sum_{r,s,\alpha}\, \lambda_\alpha\, (C_\alpha)_{r'r}\,(\rho_i)_{rs}\,(C_\alpha^\dag)_{ss'}\, \nonumber \\
& =&\sum_{\alpha} \lambda_\alpha\, \left(C_\alpha\, \rho_i\, C^\dag_\alpha\right)_{r's'} \nonumber \\
\Longrightarrow  \ \ \  \rho_f&=&\sum_{\mu}\lambda_{\mu}\, C_{\mu}\, \rho_i\,C^\dag_{\mu}. 
\end{eqnarray}   
\item From (\ref{ele}) the trace preservation condition (\ref{A2}) assumes the form 
\begin{eqnarray}
\label{tp2}
{\rm Tr}(\rho_f)=1\ \ \ &&\Longrightarrow  \sum_{\alpha}\lambda_{\alpha}\, {\rm Tr}\,\left(C_{\alpha}\, \rho_i\,C^\dag_{\alpha}\right)= 
\sum_{\alpha}\lambda_{\alpha}\, {\rm Tr}\,\left(C^\dag_{\alpha}\,C_{\alpha}\, \rho_i\,\right)=1 \nonumber \\ 
&&\Longrightarrow  \sum_{\alpha}\lambda_{\alpha}\,C^\dag_{\alpha}\,C_{\alpha}=I_{n}. 
\end{eqnarray}   
\item  From (\ref{abdef}) and (\ref{canA2}) we may identify the elements of the realigned $B$ matrix as  
\begin{equation}
\label{bac}
B_{r'r;s's}=  \sum_\alpha\, \lambda_\alpha\, (C_\alpha)_{r'r}\, (C_\alpha^*)_{s's} 
\end{equation}   
which happens to be the spectral decomposition of the dynamical $B$ matrix with  $\lambda_\mu$ being its eigenvalues.
\end{itemize}
Highlighting point here is that (\ref{bac}) brings out an explicit  connection between the hermitian (coefficient) matrix $\mathcal{A}$ (see (\ref{smalla1}) and (\ref{aher}))  and the dynamical matrix $B$ of Ref.~\cite{ECGS1}: 
 \begin{enumerate}
 \item The eigenvalues of the coefficient matrix  $\mathcal{A}$ associated with the $A$-form are identically same as those of  $B$. 
 \item A completely positive map requires that the coefficient matrix $\mathcal{A}$ is positive (i.e., the eigenvalues $\lambda_{\alpha}$ are  non-negative whenever the map is completely positive). 
 \item In the case of a completely positive map one may define a set $\{E_\alpha,\,\alpha=0,1,\ldots,n^2-1\}$ of  $n\times n$ matrices based on the canonical structure (\ref{canA}) of the $A$-map:
 \begin{equation}
 \label{krE}
 E_\alpha=\sqrt{\lambda_\alpha}\, C_\alpha.
 \end{equation}
Then the transformation $\rho_i~\rightarrow~\rho_f~=~A(\rho_i)$ gets expressed in terms of the Kraus operator-sum representation~\cite{Kraus} i.e., 
  \begin{eqnarray}
  \label{krAmap}
  (\rho_f)_{r's'}&=& \sum_{r,s}\, A_{r's';rs}\, (\rho_i)_{rs} \nonumber \\ 
  &=&  \sum_{r,s,\alpha} \, \lambda_\alpha\, (E_\alpha)_{r'r}\,(E^*_\alpha)_{s's}\,(\rho_i)_{rs} \nonumber \\ 
  &=& \sum_\alpha (E_\alpha)_{r'r}\,(\rho_i)_{rs}\, (E^\dag_\alpha)_{ss'}=\sum_\alpha\, \left(E_\alpha\,\rho_i\,E^\dag_\alpha\right)_{r's'}  \nonumber \\
  &&\Longrightarrow \rho_f=\sum_\alpha\, E_\alpha\,\rho_i\,E^\dag_\alpha.  
  \end{eqnarray}
  We point out  that the operator sum representation (\ref{krAmap}) was already described (via  the spectral decomposition of the dynamical matrix $B$)  by Sudarshan, Mathews and Rau in  their 1961 paper~\cite{ECGS1} and it was independently proposed by Kraus~\cite{Kraus} after 10 years. The operators  $E_\alpha$ (see  (\ref{krE}),(\ref{krAmap}))  associated with a completely positive map are known as Kraus operators in the literature.  
 	\end{enumerate}
 
 Summarizing, in this section we have shown that the canonical structure (\ref{canA}) of the $A$-form plays a significant role on its own --   bringing forth all the required features of the quantum channel --  without any necessity to invoke the realigned $B$-form.  In the next section  we employ the  $A$-form to elucidate the completely positive or not completely positive behaviour of some familiar qubit channels.

\section{Canonical $A$\,-\,form of standard qubit maps} 

In this section we illustrate explicit $4\times 4$ matrix forms of the canonical $A$-form and its equivalence with the dynamical matrix $B$ of some standard qubit transformations.  

\subsection{Unitary map} 
Under the action of a unitary transformation we have  
\begin{eqnarray}
\label{ss21}
\rho_f=U\,\rho_i\,U^\dagger\,\ \Longrightarrow 
\left(\rho_f\right)_{r's'}&=&\sum_{r,s}\,U_{r'r}\,U^*_{s's}\,\left(\rho_i\right)_{rs} =\sum_{r,s}\,\left(U\otimes U^*\right)_{r's'; rs}\,\left(\rho_i\right)_{rs}. 
\end{eqnarray} 
We thus obtain (see (\ref{defA})) 
\begin{equation}
\label{ss22}
\left(A_U\right)_{r's';rs}=\left(U\otimes U^*\right)_{r's'; rs} \Longrightarrow A_U=U\otimes U^* 
\end{equation}

Let us consider the $2\times 2$ unitary matrix  
\begin{eqnarray}
\label{22u}
U&=&e^{i\, (\vec{\sigma}.\hat{n})\theta/2}, \ \ \vert\hat{n}\vert^2=n_1^2+n_2^2+n_3^2=1  \nonumber \\ 
&=& I_2\, \cos\left(\theta/2\right)  + i\,  \vec{\sigma}.\hat{n}  \, \sin\left(\theta/2\right)  \nonumber \\
&=& \left(\begin{array}{cc}\cos\left(\frac{\theta}{2}\right) +i\, n_3\,\sin\left(\frac{\theta}{2}\right)  & i\,n_-\,\sin\left(\frac{\theta}{2}\right) \\   
i\,n_+\,\sin\left(\frac{\theta}{2}\right) & \cos\left(\frac{\theta}{2}\right) -i\, n_3\,\sin\left(\frac{\theta}{2}\right)  \\ 
 \end{array}\right),\ \ n_\pm=(n_1\pm i\,n_2).
\end{eqnarray}
Then the matrix $A_U$ (see (\ref{defAU}), (\ref{ss22})) associated with the unitary matrix (\ref{22u}) takes the form  
 \begin{eqnarray} 
\label{amunu}
A_U &=& \left(I_2\otimes I_2\right)\, \cos^2\left(\theta/2\right)   + \left(\vec{\sigma}^*\cdot\hat{n}\otimes\vec{\sigma}^*\cdot\hat{n}\right)\, \sin^{2}\left(\theta/2\right) \nonumber \\ 
&& \ \  -i\left(I_2\otimes\vec{\sigma}^*\cdot\hat{n}- \vec{\sigma}\cdot\hat{n}\otimes I_2\right)\, \cos\left(\theta/2\right)  
\sin\left(\theta/2\right). 
\end{eqnarray}
Denoting  $\sigma_0=I_2$ and employing the basis set   $\{T_\mu=\frac{1}{\sqrt{2}}\, \sigma_\mu\}$ of $2\times 2$ matrices satisfying  (see (\ref{tor})) the conditions 
$$\frac{1}{2}{\rm Tr}[\sigma_\mu\,\sigma_\nu]=\delta_{\mu\nu},\ \mu,\nu=0,1,2,3,$$ we express (\ref{amunu}) in the following compact form. 
\begin{eqnarray}
\label{bau22}
A_U=\frac{1}{2}\,\sum_{\mu,\nu=0,1,2,3}\, (a_U)_{\mu\nu}\, \sigma_{\mu}\otimes \sigma^*_{\nu}, 
\end{eqnarray}
Observe that 
\begin{eqnarray}
\label{amunu2}
(a_U)_{\mu\nu}&=&\frac{1}{2}\,{\rm Tr}\left[A_U\,(\sigma_\mu\otimes\sigma^*_\nu)\right] \nonumber \\ 
&=& \frac{1}{2}\,{\rm Tr}\left[U\,\sigma_\mu\right]\, {\rm Tr}\left[U\,\sigma_\nu\right]^*
\end{eqnarray}
where ${\rm Tr}\left[U\,\sigma_\mu\right],\mu=0,1,2,3$ is evaluated using (\ref{22u}):     
\begin{eqnarray}
{\rm Tr}\left[U\,\sigma_0\right]= 2\,\cos\left(\frac{\theta}{2}\right), \ \  
{\rm Tr}\left[U\,\sigma_k\right]= 2i\,n_k\, \sin\left(\frac{\theta}{2}\right), \ \  k=1,2,3  
\end{eqnarray}
Then the $4\times 4$ coefficient matrix $\mathcal{A}_U=((a_U)_{\mu\nu})$ (see (\ref{amunu2})) associated with $A_U$  is given by 
\begin{equation}
\label{coau}
\mathcal{A}_U= 2\,  X_U\, X_U^\dag,  \ \ \  
X_U=\left(\begin{array}{c} \cos\left(\frac{\theta}{2}\right) \\ i\,n_1\, \sin\left(\frac{\theta}{2}\right) \\ 
i\,n_2\, \sin\left(\frac{\theta}{2}\right) \\ i\,n_3\, \sin\left(\frac{\theta}{2}\right) \end{array}\right). 
\end{equation}
From (\ref{coau}) it is seen that the coefficient matrix $\mathcal{A}$ is a rank-1 positive matrix with  eigenvalue 2 and eigenvector $X_U$. 
The realigned $B_U$ matrix matches exactly with the coefficient matrix $\mathcal{A}_U$ (see (\ref{coau})) i.e.,  
\begin{equation}
B_U=2\, X_U X_U^\dag \equiv \mathcal{A}_U.
\end{equation}

\subsection{Pin map: } 

Consider a linear $A$-form mapping every input state $\rho_i$ to a fixed output state $\rho_0$   i.e., 
\begin{equation}
A_{\rm pin}: \rho_{i}\rightarrow \rho_0=A_{\rm pin}(\rho_i) \ \forall\rho_i. 
\end{equation}  
Sudarshan, Mathews and Rau presented this map in terms of the $B$-form, which was termed as  {\em relaxation generator}. Here we would like to illustrate the canonical structure of the $A$-form associated with the qubit pin-map. 

Let the fixed output density matrix of the qubit be given by  
\begin{equation} 
\label{pinrho}
\rho_0=\frac{1}{2}\left(I_2+ \vec\sigma\cdot \vec{p}_0 \right).  
\end{equation} 
The $4\times 4$ matrix  $A_{\rm pin}$ corresponding to the pin map is identified as follows:   
\begin{eqnarray}
\label{pinex} 
\left(\begin{array}{c} 1+\left(p_0\right)_{3} \\ \left(p_0\right)_{1}-i \left(p_0\right)_{2}\\ \left(p_0\right)_{1}+i \left(p_0\right)_{2}  \\ 1-\left(p_0\right)_{3} \end{array}\right) &=& 
\left(\begin{array}{cccc}  \left(A_{\rm pin}\right)_{00;00} & \left(A_{\rm pin}\right)_{00;01} & \left(A_{\rm pin}\right)_{00;10} & \left(A_{\rm pin}\right)_{00;11}  \\  
\left(A_{\rm pin}\right)_{01;00} & \left(A_{\rm pin}\right)_{01;01} & \left(A_{\rm pin}\right)_{01;10} & A_{01;11}  \\
\left(A_{\rm pin}\right)_{10;00} & \left(A_{\rm pin}\right)_{10;01} & \left(A_{\rm pin}\right)_{10;10} & \left(A_{\rm pin}\right)_{10;11}  \\
\left(A_{\rm pin}\right)_{11;00} & \left(A_{\rm pin}\right)_{11;01} & \left(A_{\rm pin}\right)_{11;10} & \left(A_{\rm pin}\right)_{11;11}  
\end{array}\right)   
\left(\begin{array}{c} 1+\left(p_i\right)_{3} \\ \left(p_i\right)_{1}-i \left(p_i\right)_{2}\\ \left(p_i\right)_{1}+i \left(p_i\right)_{2}  \\ 1-\left(p_i\right)_{3} \end{array}\right)\ \ \forall \ \ \vec{p}_i \nonumber \\ 
&& \Longrightarrow  A_{\rm pin}=\frac{1}{2}\,\left(\begin{array}{cccc}  1+\left(p_0\right)_3 & 0 & 0 & 1+\left(p_0\right)_3  \\  
\left(p_0\right)_1-i\left(p_0\right)_2 & 0 & 0 & \left(p_0\right)_1-i\left(p_0\right)_2  \\
\left(p_0\right)_1+i\left(p_0\right)_2 & 0 & 0 & \left(p_0\right)_1+i\left(p_0\right)_2  \\
1-\left(p_0\right)_3 & 0 & 0 & 1-\left(p_0\right)_3  
\end{array}\right).
\end{eqnarray}   
Using the orthonormal basis set of  matrices $\{\frac{\sigma_\mu}{\sqrt{2}}, \mu=0,1,2,3\}$ we expand 
$$A_{\rm pin}=\frac{1}{2}\sum_{\mu,\nu}\, \left(a_{\rm pin}\right)_{\mu\nu}\, \sigma_{\mu}\otimes\sigma_\nu^*,\ {\rm where}\  \left(a_{\rm pin}\right)_{\mu\nu}=
\frac{1}{2}{\rm Tr}[A_{\rm pin}\,\left(\sigma_\mu\otimes\sigma_\nu^*\right)].$$ 
 The coefficient matrix $\mathcal{A}_{\rm pin}=(\left(a_{\rm pin}\right)_{\mu\nu})$ is then found to be  
\begin{eqnarray}
 \mathcal{A}_{\rm pin}=\frac{1}{2}\left(\begin{array}{cccc}  1& \left(p_0\right)_1  & \left(p_0\right)_2  & \left(p_0\right)_3  \\  
 	\left(p_0\right)_1 & 1 & -i\left(p_0\right)_3 & i\left(p_0\right)_2  \\
 	\left(p_0\right)_2 & i\left(p_0\right)_3 & 1 & -i\left(p_0\right)_1  \\
 	\left(p_0\right)_3 & -i\left(p_0\right)_2 & i\left(p_0\right)_1 & 1  
 \end{array}\right).
\end{eqnarray}   
The eigenvalues of $\mathcal{A}_{\rm pin}$ are given by   
\begin{equation}
\label{evpin}
\lambda_0=\lambda_1=\frac{1}{2}\, (1+\vert\vec{p}_0\vert), \ \  \lambda_2=\lambda_3=\frac{1}{2}(1-\vert\vec{p}_0\vert). 
\end{equation} 
Clearly, the eigenvalues of $\mathcal{A}_{\rm pin}$ are all positive ensuring that the pin map is completely positive.   

 The $B$ matrix associated with the pin map is constructed using the explicit matrix form of $A_{\rm pin}$ (see (\ref{pinex})):   
\begin{eqnarray}
	B_{\rm pin}&=&\frac{1}{2}\left(\begin{array}{cccc} 1+\left(p_0\right)_3 & 0 & \left(p_0\right)_1-i\,\left(p_0\right)_2 & 0 \\ 0 & 1+\left(p_0\right)_3 & 0 & \left(p_0\right)_1-i\, \left(p_0\right)_2 \\ 
		\left(p_0\right)_1+i\, \left(p_0\right)_2 & 0 & 1-\left(p_0\right)_3 & 0 \\ 
		0 & \left(p_0\right)_1+i\, \left(p_0\right)_2& 0 & 1-\left(p_0\right)_3 \end{array}\right)\nonumber \\ 
		&=&\rho_0\otimes I_2 
		\end{eqnarray} 
Eigenvalues of $B_{\rm pin}$ match with those of the coefficient matrix $\mathcal{A}_{\rm pin}$ (see (\ref{evpin})), thus establishing the equivalence between the two.

\subsection{Transpose map:}
 Consider the transpose map $A_T:\rho\rightarrow\rho^T$ on  qubit density matrices (see (\ref{qrho})).  
We obtain the associated $4\times 4$ matrix form of $A_T$ as  
\begin{equation}
\label{at}
A_T=
\left(\begin{array}{cccc}\,1 & 0 & 0 & 0 \\ 
0 & 0 & 1 & 0\\ 0 & 1 & 0 &  0 \\ 0 & 0 & 0 & 1 \\ 
\end{array}\right). 
\end{equation} 
Employing the  basis set  $\{\frac{\sigma_\mu}{\sqrt{2}}, \mu=0,1,2,3\}$ we express 
\begin{equation}
A_T=\frac{1}{2}\sum_{\mu,\nu=0}^3\, \left(a_{T}\right)_{\mu\nu}\, \left(\sigma_{\mu}\otimes \sigma^*_{\nu} \right)
\end{equation}
to obtain the following explicit structure for the coefficient matrix $\mathcal{A}_T$: 
\begin{equation}
\label{calat}
\mathcal{A}_T=\mbox{diag}\,\left(-1,\,1,\,1,\,1 \right). 
\end{equation}
The matrix $\mathcal{A}_T$ is not positive (one of the eigenvalues of $\mathcal{A}_T$ is -1) which points towards the not-completely positive nature of the transpose map.  

From the explicit matrix structure of  $A_T$ (see (\ref{at})) it is easy to see that the realigned dynamical matrix  $B_T\equiv A_T$. The 
eigenvalues of $B_T$ match with those of the coefficent matrix $\mathcal{A}_T$ (see (\ref{calat})). 

\subsection{Projection of the Bloch sphere onto its equatorial plane:} 
A map that projects the entire Bloch sphere onto the equatorial plane is defined by the transformation of the Bloch vector  
 $$(p_1,\,p_2,\,p_3)\rightarrow (p_1,\,p_2,\,0).$$ This leads to the following linear transformation    
\begin{equation}
\label{pR}
\left(\begin{array}{c}\,\ 1 \\  p_1-i\ p_2 \\ p_1+i\ p_2 \\ 1 \end{array}\right)  =
\frac{1}{2}\left(\begin{array}{cccc}
1 & 0 & 0 & 1 \\ 
0 & 2 & 0 & 0\\ 
0 & 0 & 2 &  0 \\ 
1 & 0 & 0 & 1  
\end{array}\right) \left(\begin{array}{c}\,1+p_3 \\  p_1-i\ p_2 \\ p_1+i\ p_2 \\ 1-p_3 \end{array}\right).  
\end{equation}
We then  express $A_P=\frac{1}{2}\sum_{\mu,\nu}\,\left(a_P\right)_{\mu\nu}\, \left(\sigma_{\mu}\otimes\sigma_\nu^*\right)$ to obtain  
\begin{equation}  
\mathcal{A}_P=\frac{1}{2}\mbox{diag}\,\left(3,\,1,\,
1,\,-1 \right).
\end{equation} 
Negative eigenvalues of $\mathcal{A}_P$ clearly indicate that projection of the Bloch sphere onto the equatorial plane is not physical as it corresponds to a not completely positive map.  The dynamical matrix $B_P$ is then obtained using the realignment   $(B_P)_{r'r;s's}=(A_P)_{r's';rs}$:    
\begin{equation}
B_P=\frac{1}{2}\left(\begin{array}{cccc}\, 1 & 0 & 0 & 2 \\ 
0 & 1 & 0 & 0\\ 0 & 0 & 1 &  0 \\ 2 & 0 & 0 & 1 \\ 
\end{array}\right).
\end{equation}
Eigenvalues  of $B_P$ are same as those of $\mathcal{A}_P$.  

	\subsection{ Bit flip channel}
		A qubit bit flip channel reverses the state of a qubit from $\vert 0\rangle$ to $\vert 1\rangle$ with probability $1-p$, $0\leq p \leq 1$; the channel keeps the states unaltered  with a probability $p$. This is a completely positive map equipped with the  Kraus operators given by~\cite{NC} 
	\begin{equation}
	\label{bf1}
	E_0=\sqrt{p}\left(\begin{array}{cc} 1 & 0 \\ 0 & 1  \end{array}\right), \ \ \ 
	E_1\equiv \sqrt{1-p}\,\left(\begin{array}{cc} 0 & 1 \\ 1 & 0  \end{array}\right).  
	\end{equation}
Using the operator-sum representation  $\rho_f= E_0\,\rho_i E_0^\dagger+
	E_1\,\rho_i E_1^\dagger$  we construct the associated $A$ matrix: 
	\begin{eqnarray} 
	\label{bitflipA}
A_{\rm BF}	&=&\left(\begin{array}{cccc}\,p & 0 & 0 & 1-p \\ 
	0 & p & 1-p & 0\\ 0 & 1-p & p & 0 \\ 1-p & 0 & 0 & p  \end{array}\right) 
	\end{eqnarray} 
	Adopting the  matrix basis $\left\{\frac{\sigma_\mu}{\sqrt{2}},\mu=0,1,2,3\right\}$, as  in all other examples studied earlier, we compute the coefficient matrix $\mathcal{A}_{\rm BF}=\left(a_{\mu\nu}\right)$ associated with  $A_{\rm BF}$:    
\begin{eqnarray}
	\mathcal{A}_{\rm BF}&=&\left(\begin{array}{cccc} 2p & 0 & 0 & 0 \\ 0 & 2(1-p) & 0 & 0 \\ 0 & 0 & 0 & 0 \\
	0 & 0 & 0 & 0 \end{array}\right)
	\end{eqnarray}
Note that the  eigenvalues $2p, 2(1-p)$ of   $\mathcal{A}_{\rm BF}$    are  positive and ascertain the completely positive nature of the bit flip channel.    
	
\subsection{Phase flip channel}
 A qubit phase flip channel is equipped with the Kraus operators~\cite{NC}   
	\begin{equation}
	\label{pf1}
	E_0=\sqrt{p}\left(\begin{array}{cc} 1 & 0 \\ 0 & 1  \end{array}\right), \ \ \ 
	E_1=\sqrt{1-p}\,\left(\begin{array}{cc} 1 & 0 \\ 0 & -1  \end{array}\right).  
	\end{equation} 
The linear map $A_{\rm PF}$ associated with the phase flip channel  is given explicitly as a $4\times 4$ matrix form:   	
\begin{eqnarray} 
	\label{pflipA}
	A_{\rm BF}&=&\frac{1}{2}\left(\begin{array}{cccc}\,1 & 0 & 0 & 0 \\ 
	0 & 2p-1 & 0 & 0\\ 0 & 0 & 2p-1 & 0 \\ 0 & 0 & 0 & 
	1 \end{array}\right).
\end{eqnarray} 
	Then the associated coefficent matrix $\mathcal{A}_{\rm BF}$ is found to be
	\begin{eqnarray}
	\mathcal{A}_{\rm BF}&=&\left(\begin{array}{cccc} 2p & 0 & 0 & 0 \\ 0 & 0 & 0 & 0 \\ 0 & 0 & 0 & 0 \\
	0 & 0 & 0 & 2(1-p) \end{array}\right)
	\end{eqnarray} 
The eigenvalues  $2p$, $2(1-p)$ of 	$\mathcal{A}_{\rm BF}$ are necessarily positive and  confirm the legitimacy (complete positivity) of the phase flip channel.  

\section{Summary}

The dynamical $A$ and $B$-maps were pioneered by Sudarshan, Mathews and Rau~\cite{ECGS1} in the context of open system dynamics. Unfortunately this seminal 1961 paper did not receive its due attention in the field, although it contained all the details of finite dimensional quantum channels.
In this work we have elaborated on the canonical structure of the $A$-form, establishing that it  offers an alternative approach to recognize the completely positive/not completely positive nature of  quantum channels. We have illustrated the canonical structure of the $A$ map in several standard examples of qubit maps.  A new geometrical representation based on the Lorentz singular value decomposition~\cite{svd} of the canonical $A$-form associated with qubit transformations is being prepared~\cite{KSAR} and it will be presented separately as a sequel to the present work.   

We dedicate this tutorial article as a mark of our reverence to Professor ECG Sudarshan.    
% We have centered our discussion on a specific choice of NCP dynamical map resulting  from a unitary evolution~\cite{Jordan} on  initially correlated two-qubit states. 

\section*{Acknowledgement}  ARU,BNK and Sudha acknowledge financial support from the Department of Science and Technology, India (Project No. DST/ICPS/QuST/Theme-2/Q107/2019).


\begin{thebibliography}{0}
	%\bibitem{Choi} M. D. Choi, Can. J. Math. {\bf 24}, 520 (1972); Linear Algebra and Appl. {\bf 10}, 285 (1975). 
	%\bibitem{Breuer} H.-P. Breuer and F. Petruccione, {\em The Theory of Open
	%Quantum Systems} (Oxford Univ. Press, Oxford, 2007).
	%
	
	\bibitem{ECGS1} Sudarshan ECG, Mathews PM, Rau J. Stochastic
	dynamics of quantum-mechanical systems. {\em Physical
		Review} 1961; {\bf 121} (3): 920–924. doi:10.1103/PhysRev.121.920
	\bibitem{Jordan1} Jordan TF, Sudarshan ECG. Dynamical mappings
	of density operators in quantum mechanics.
	{\em Journal of Mathematical Physics} 1961; {\bf 2} (6):
	772–775. doi:10.1063/1.1724221
	\bibitem{NC} M. A. Nielsen MA,  Chaung IL. {\em  Quantum Computation and
		Quantum Information}. Cambridge: Cambridge University Press,
	2002. 
	\bibitem{Breuer} Breuer H-P, Petruccione F. {\em The Theory of Open
		Quantum Systems}. Oxford: Oxford University Press, 2007.
	\bibitem{simon}  Simon S,  Rajagopalan SP,  Simon R. The structure of states and maps in quantum theory. {\em Pramana - Journal of Physics} 2009; {\bf 73} (3):
	471-483. https://www.ias.ac.in/article/fulltext/pram/073/03/0471-0483 
	\bibitem{Vinayak} Jagadish V, Petruccione F. An Invitation to Quantum
	Channels. {\em Quanta} 2018; {\bf 7} (1): 54-57. doi:10.12743/quanta.v7i1.77 
	\bibitem{urs} Usha Devi AR, Rajagopal AK, Sudha.  Open-system quantum dynamics with correlated initial states, not completely positive maps,
	and non-Markovianity. {\em Physical Review A} 2011;  {\bf 83}(2): 022109. doi: 10.1103/PhysRevA.83.022109. 
	\bibitem{Kraus} Kraus K. General state changes in quantum theory.
	{\em Annals of Physics} 1971; {\bf 64}(2): 311–335. doi:10.1016/0003-4916(71)90108-4
\bibitem{svd} Sudha,  Karthik HS, Pal R, Akhilesh KS, Ghosh S, Mallesh KS,  Usha Devi, AR. Canonical forms of two-qubit states under local operations. {\em Physical Review A} 2020; {\bf 102}(5): 052419. doi: 10.1103/PhysRevA.102.052419. 
	\bibitem{KSAR} Karthik BN, Sudha, Usha Devi AR, Rajagopal AK. Geometrical representation of canonical $A$-form for qubit transformations. Under preparation.    
	%\bibitem{Pechukas} P. Pechukas, Phys.Rev.Lett. {\bf 73}, 1060 (1994).
	%\bibitem{Alicki} R. Alicki, Phys.Rev.Lett {\bf 75}, 3020 (1995); 
	%\bibitem{Pechukas1995} P. Pechukas, Phys.Rev.Lett. {\bf 75}, 3021 (1995). 
	%\bibitem{Buzek} P. Stelmachovic and V. Buzek, Phys.Rev.A {\bf 64}, 062106(2001).
	%\bibitem{Terno} H. A. Carteret, D. R. Terno, and K. {\. Z}yczkowski, Phys.Rev.A. {\bf 77}, 042113 (2008). 
	%\bibitem{Lidar} A. Shabani and D. A, Lidar, Phys.Rev.Lett. {\bf 102}, 100402 (2009). 
	%\bibitem{Jordan} T. F. Jordan, A. Shaji, and E. C. G. Sudarshan, Phys.Rev.A. {\bf 70}, 052110 (2004).
	%\bibitem{Rosario} C. A. Rodr{\' i}guez-Rosario, K. Modi,  Aik-meng Kuah, A. Shaji, and E. C. G. Sudarshan, J. Phys. A {\bf 41}, 205301 (2008). 
	%\bibitem{ECGS} C. A. Rodr{\' i}guez-Rosario and E. C. G. Sudarshan, arXiv:0803.1183 (quant-ph).
	%\bibitem{Modi} K. Modi and E. C. G. Sudarshan, Phys.Rev.A {\bf 81}, 052119 (2010).  
	% 
	
	%\bibitem{Zurek} H. Ollivier and W. H. Zurek, \prl {\bf 88}, 017901 (2001).
	
	%\bibitem{Lindblad} G. Lindblad, Comm. Math. Phy. {\bf 48}, 119 (1976). 
	%\bibitem{GKS} V. Gorini, A. Kossakowski, and E. C. G. Sudarshan, J. Math. Phys. {\bf 17}, 821 (1976).
	% \bibitem{Daffer} S. Daffer, K. W{\' o}dkiewicz, J. D. Cresser, and J. K. McIver \pra {\bf 70}, 5010304 (2004).
	%\bibitem{Breuer2}  H.-P. Breuer, \pra {\bf 70}, 012106 (2004).
	%\bibitem{Kossakowski} D. Chrus{\' c}in{\' s}ki, and A. Kossakowski, \prl {\bf 104}, 070406 (2010); D. Chrus{\' c}in{\'s}ki, and A. Kossakowski and 
	%S. Pascazio, \pra {\bf 81}, 032101 (2010).
	%\bibitem{Cirac} M.M. Wolf, J. Eisert, T.S. Cubitt, and J.I. Cirac, \prl {\bf 101}, 150402 (2008).
	%\bibitem{B3} H.-P. Breuer et al. , E.-M. Laine, J. Piilo, Phys. Rev.
	%Lett. 103, 210401 (2009); E.-M. Laine, J. Piilo, H.-P. Breuer, \pra {\bf 81}, 062115 (2010).
	%\bibitem{Angel} A. Rivas, S. F. Hulega, M. B. Plenio, \prl {\bf 105}, 050403 (2010).
	%\bibitem{AKRU} A. K. Rajagopal, A. R. Usha Devi, R. W. Rendell, \pra {\bf 82}, 042107 (2010).  
%	\bibitem{kns} K. N. Srinivasa Rao, {\emph{Rotation and Lorentz Groups for Physicists}}, (Wiley, 1988)
%	\bibitem{Choi} M. D. Choi, Can. J. Math. {\bf 24}, 520 (1972); Linear Algebra and Appl. {\bf 10}, 285 (1975).  
	%\bibitem{note} The eigenvalues of ${\cal A}$ are the same as that of the $B$-matrix considered in Ref.~\cite{Jordan}.  
	%\bibitem{NC} M. A. Nielsen and I. L. Chaung, {\em Quantum computation and quantum information} (Cambridge Univ. Press, Cambridge, 2002).  
	%\bibitem{MBR} M. B. Ruskai, J. Math. Phys. {\bf 43}, 4358 (2002).
	%\bibitem{Jozsa} R. Jozsa, J. Mod. Optics, 41, 2315 (1994). 
%	\bibitem{nc} M. A. Nielsen and I. L. Chaung, {\emph {Quantum Computation and
%			Quantum Information}} (Cambridge University Press, Cambridge, 2002).
%	\bibitem{Kraus} K. Kraus, States, Effects and Operations: Fundamental
%	Notions of Quantum Theory, vol. 190 of Lecture notes in Physics (Spring-Verlag, New York, 1983). 
	
\end{thebibliography}
\end{document}